\documentstyle[graphicx,epsf]{basi}
\begin{document}
\title{Study of Bubble Nebula using IUE high resolution Spectra}
\author[Anand et al.]
{Anand M.Y.$^{1}$, Kagali B.A.$^1$ \& Jayant Murthy$^2$\\
$^1$Department of Physics, Bangalore University, Bangalore 560056,India\\
$^2$ Indian Institute of Astrophysics, Bangalore - 560034\\}
\maketitle
\label{firstpage}
\begin{abstract}
In this paper we have analyzed IUE high resolution spectra of the central star (BD+602522) of the Bubble nebula. We discuss 
velocities of the different regions along the line of sight to the bubble. We find that the Bubble Nebula is younger (by a factor 
of 100) than the exciting star suggesting that either the bubble is expanding into an inhomogenuous interstellar medium or that 
the mechanics of the stellar wind are not fully understood.
\end{abstract}



\section{Introduction}
\label{sec:intro}
Interstellar bubbles are formed by the interaction of stellar winds (or supernovae) with the interstellar medium. As the material 
from the star streams out at supersonic velocities, it encounters the ambient ISM forming a shock front. This interaction is one 
of the main mechanisms by which energy is transferred from stars to the ISM and maintains the hot phase of the three phase ISM 
(McKee and Ostriker 1977). 
The first model of interstellar bubbles was created by Weaver (1977) who found that four regions would form around a hot star with 
a strong stellar wind. Immediately around the star is a volume comprised essentially of the stellar wind, from which the 
interstellar matter has been swept clean. A shock forms where the supersonic stellar wind runs into the interstellar medium 
followed by a region of swept up interstellar matter. The ambient ISM, which has not yet experienced the shock, surrounds the 
entire bubble. This picture is illustrated in Fig. 1. 

\begin{figure}
\centering
\includegraphics*[width=2.3 in,] {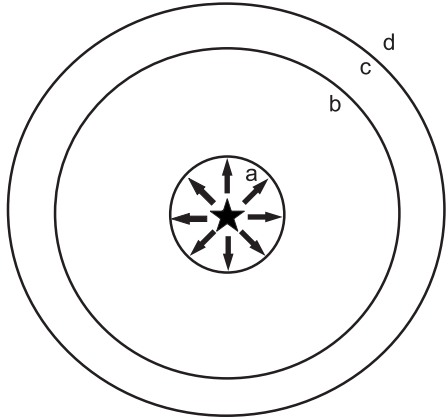}
\caption{Structure of bubble as given by Weaver et al. model. a: Freely expanding supersonic stellar wind region. b: Shocked 
stellar wind region c: Swept up ISM d: Ambient ISM}
\label{fig:1}
\end{figure}

Although this simple model is appealing and explains the gross features of interstellar bubbles, Naze et al. (2001) have pointed 
out discrepancies between the dynamical time scales of the bubbles in N11B H II region in Large Magellanic Cloud which are much 
shorter than the age of the OB association as well as inconsistencies between the observed and predicted luminosities of the 
bubbles. Similar discrepancies have been noted in superbubbles (Oey 1996).  We have chosen to investigate a well-studied 
interstellar bubble in order to further test the standard model for bubbles.

 The Bubble Nebula (NGC 7635) is a prominent object in the complex of interlocking shells belonging to the emission nebula S162, 
situated near the Galactic plane in the Perseus arm at a distance of about 3.6 kpc (Table 1). As the name suggests, it is a 
spherical nebula energized by a single, off-center O star (BD +60 25 22). The physical properties of this star are summarized in 
Table 1, taken from Christopoulou et al. (1995). 
The Bubble nebula, itself, has been mapped in the entire spectral band from the optical to the infrared and radio. Based on direct 
photographic and Fabry-perot interferogram techniques, Pismis et al(1983) have found an age of 3-4$\times 10^{5}$ years for the 
bubble. The shell velocity was estimated to be 35 km/s by Christopoulou through H$\alpha$  and NII emission line studies. From 
this, they have estimated a dynamical lifetime of only 5$\times 10^{4}$ years for this high velocity shell suggesting that this 
expansion is a relatively short lived phenomenon. The infrared luminosity of the bubble is much less (by a factor of 4) than that 
of the central star with a low abundance of $^{12}{\rm CO}$ (Thronson et al 1982).

\begin{table*}
\caption{The relevant Astronomical data}
\begin{tabular} {l l l}
\hline
parameters &     &    Reference\\[0.5ex]
\hline
BD number                 &   60$^{\circ}$ 2522        &                         \\
Galactic longitude (l)    &   112$^{\circ}$.3          &                         \\
Galactic latitude (b)     &   0$^{\circ}$.2            &                         \\ 
R.A.                      &   23 20 44.2               &    ep-2000              \\
Declination               &   +61 11 40.6              &    ep-2000              \\                        
Spectral type             &   O6.5III ef               &  Conti \& Leep 1974     \\ 
E(B-V)                    &   0.73                                               \\
Distance( kpc)            &   3.6             &  Johnson 1982                                \\
Terminal wind velocity(km/s) & 1800           &  Johnson 1980                    \\
                             & 2500           &  Leitherer 1988                  \\
Rate of stellar mass loss    &                &                                  \\
$({\rm M}_{\odot} {\rm yr}^{-1}$)  & ${\rm 10}^{-5.67}$     &  Leitherer 1988          \\[1ex]
\hline
\end{tabular}
\label{table:nonlin}
\end{table*}

\begin{figure}
\centering
\includegraphics*[width=3.1 in,]{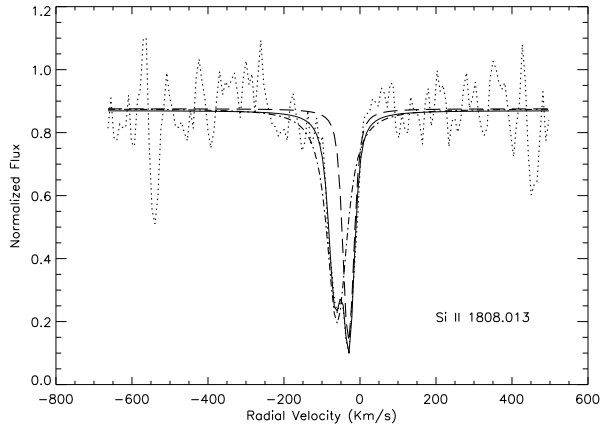}
\caption{Plot of the fitted profile to the Si II 1808.0126 $\AA$ line. The dotted line represent observed line profile and solid 
lines represent fitted profile}
\end{figure}

\section {IUE Archival Data}
\begin{table*}
\caption{IUE observation of BD+602522}
\centering
\begin{tabular} {cccc}
\hline
Image & date & Exposure time(sec)& Aperture  \\[0.5ex]
\hline
SWP 08840   & 1980 April 27   & 10199.729 &	  L  \\
            &                 &           &      \\
LWR 07625   &	1980 April 27 & 8999.605  &	  L  \\
            &                 &           &      \\
\hline
\end{tabular}
\label{table:nonlin}
\end{table*}

We have used archival data from the International Ultraviolet Explorer (IUE) to study the Bubble Nebula. IUE was launched in 1977 
for a nominal 3 year mission and was finally shut off in 1999 after 18 years of operation. The instrument consists of a telescope, 
two ultraviolet spectrographs and four cameras: SWP (short wavelength prime); SWR (short wavelength redundant); LWP (long 
wavelength prime) and LWR (long wavelength redundant). The short wavelength cameras operate in the spectral range 1150-2100 $\AA$ 
while the long wavelength cameras operate in the spectral range between 1845-2980 $\AA$. Each camera is operable with either a low 
resolution (R = 200) spectrograph or an echelle where R is 10,000 for SWP/R and 15,000 for LWP/R. Details of the IUE 
instrumentation and operation are given by Boggess et al (1978). 

We have searched through the IUE archives at the Space Telescope Science Institute for high dispersion observations of the central 
star of the Bubble Nebula and have found two images: LWR07625 and SWP08840. Details of these two observations are given in Table 2 
and the data have been downloaded as fully calibrated spectra. 
The entire IUE archive has been reprocessed using the New Spectral Image Processing System (NEWSIPS) which provides a uniform set 
of spectra incorporating the best available improvements in the reduction algorithms and calibration. A full description of the 
NEWSIPS used to process IUE data and of the archived data products is given by Nichols and Linsky (1996). 

A complete list of lines observed in the interstellar medium has been given by Morton (1991) with their central wavelengths and 
oscillator strengths. We have fit the spectrum around each of the interstellar lines using a Voigt profile with the central 
wavelength, column density, velocity parameter (b), and continuum level as free parameters. The resultant profile was convolved 
with a Gaussian with a FWHM defined by $\lambda/\bigtriangleup\lambda$ =10,000 and 15,000 for the SWP and LWR spectra, 
respectively and compared with the observed spectra to yield a $\chi^{2}$ value. We then used the method of Lampton, Margon \& 
Bowyer (1976) to set limits on each of the model parameters. This consists of increasing each parameter in turn while keeping the 
other parameter fixed untill the reduced chisq exceed the required confidence interval.

\begin{figure}
\centering
\includegraphics*[width=3.1 in,]{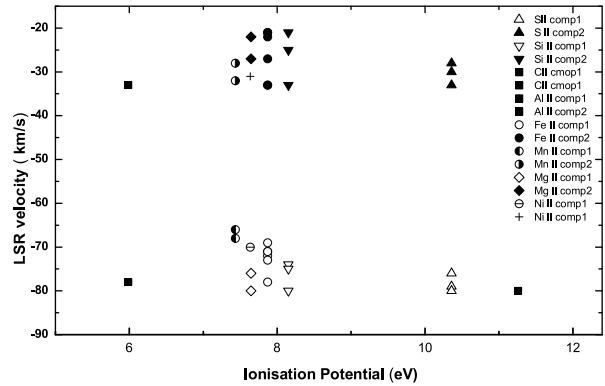}
\caption{Plot of LSR velocities of the two components.}
\end{figure}

\begin{figure}
\centering
\includegraphics*[width=3.1 in,]{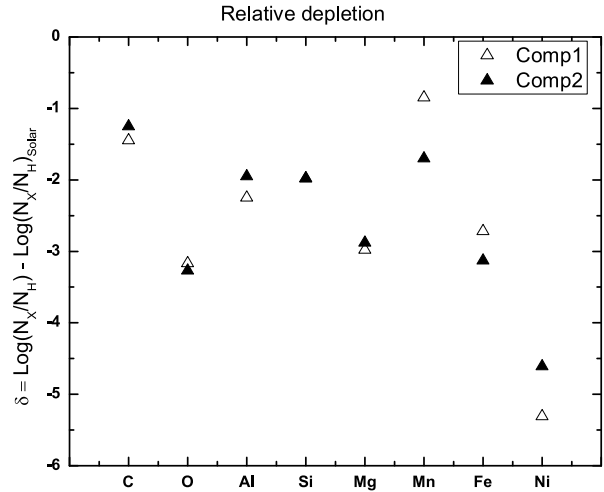}
\caption{Plot of depletion along the line of sight.}
\end{figure}

\section{Results} 
A complete list of the lines observed in our spectra is in Table 3 with their velocities in the local standard of rest (LSR) and 
their equivalent widths. We first fit the lines with a single component Voigt profile but found that, in most cases, two 
components were actually present (Fig. 2) with distinct velocities (Fig. 3). One of these components (Component 1) has a velocity 
about order of 60 km/s characteristic of that found in the surrounding shell in the radio (45 km/s, Thronson et al., 1982) and we 
have identified it with material from the ISM that has been swept up by the expansion of the bubble. The column density in 
Component 1 is about $1.5 \times 10^{21}  {\rm cm}^{-2}$  from our IUE analysis, consistent with the $2 - 5 \times 10^{21}  {\rm 
cm}^{-2}$ (Thronson et al. 1982) in the surrounding nebula, and its velocity with respect to the exciting star is about 40 km/s. 
We find no evidence for this ambient medium in our spectra perhaps indicating that most of the material has already been swept up 
and that the bubble will soon break out into the low density gas around.

The other component (Component 2) is likely to be from the intervening interstellar medium and not associated with the Nebula. An 
examination of the line velocities in IUE spectra along the nearby lines of sight (Table 4) to AR Cas (176 pc) and 1 Cas (338 pc) 
yield a similar column density and velocity to Component 2 consistent with its origin in an interstellar cloud within about 170 pc 
from the Sun. Fig 5 and 6 shows curve of growth plotted for component 1 and 2 respectively. Fig 4 represents the depletion found 
from these two components.

The remaining two sets of lines (CIV \& Al III) in Table 3 are characteristic of hotter regions and are likely to be formed in 
either the conductive interface (C IV) or in the HII region around the nebula (Al III). Their velocities with respect to the 
central star are on the order of 20 - 30 km/s.

We will compare our observational results with those expected from the model in the next section. 
\begin{table*}
\begin{center} {\footnotesize}
\caption{LSR velocities of the two components along the line of sight with 1$\sigma$ errors.}
\begin{tabular}{cccccc}
\hline
& & \multicolumn{2}{c}{$Component 1$} & \multicolumn{2}{c}{$Component 2$} \\
elements & lab wavelength & LSR velocity & eqwdth   & LSR velocity & eqwdth  \\
         &  (\AA)         & (km/s)       & ({m\AA}) &  (km/s)      & ({m\AA})  \\
\hline

C IV      &   1548.195  & $-63\pm 13$  & $253\pm 20$  &     -       &    -         \\
          &   1550.790  & $-63\pm 15$  & $92\pm 22$   &     -       &    -         \\
Al III    &   1854.716  & $-73\pm 8$   & $123\pm 17$  &     -       &    -         \\
          &   1862.790  & $-78\pm 5$   & $218\pm 27$  &     -       &    -         \\
S II      &   1250.584  & $-74\pm 10$  & $163\pm 10$  & $-30\pm 3$  & $50\pm 20  $ \\
          &   1253.811  & $-79\pm 9$   & $120\pm 44$  & $-22\pm 10$ & $39\pm 13  $ \\
          &   1259.519  & $-81\pm 14$  & $53\pm 5$    & $-36\pm 4$  & $103\pm 12 $ \\
O I       &   1302.169  & $-80\pm 15$  & $137\pm 7$   & $-14\pm 3$  & $97\pm 17  $ \\
C II      &   1334.532  & $-81\pm 14$  & $257\pm 10$  & $-23\pm 5$  & $180\pm 16 $ \\
Si II     &   1304.370  & $-79\pm 11$  & $114\pm 27$  & $-25\pm 8$  & $69\pm 18  $ \\
          &   1526.707  & $-74\pm 12$  & $382\pm 17$  & $-20\pm 9$  & $152\pm 21 $ \\
          &   1808.013  & $-75\pm 7$   & $264\pm 21$  & $-36\pm 3$  & $158\pm 12 $ \\
C I       &   1560.309  & $-79\pm 9$   & $159\pm 17$  & $-28\pm 2$  & $114\pm 7  $ \\
Al II     &   1670.787  & $-78\pm 10$  & $343\pm 67$  & $-33\pm 4$  & $277\pm 11 $ \\
Fe II     &   1608.451  & $-81\pm 6$   & $118\pm 32$  & $-33\pm 3$  & $254\pm 20 $ \\
          &   2344.214  & $-64\pm 8$   & $154\pm 50$  & $-14\pm 4$  & $472\pm 33 $ \\
		  &   2374.461  & $-71\pm 4$   & $228\pm 33$  & $-36\pm 3$  & $226\pm 27 $ \\
		  &   2382.765  & $-72\pm 14$  & $366\pm 72$  & $-15\pm 4$  & $370\pm 107$ \\
		  &   2586.650  & $-71\pm 3$   & $89\pm 21$   & $-27\pm 3$  & $304\pm 36 $ \\
 		  &   2600.173  & $-66\pm 3$   & $283\pm 33$  & $-12\pm 6$  & $530\pm 65 $ \\		
Ni II     &   1741.549  & $-70\pm 3$   & $83\pm 14$   & $-31\pm 3$  & $73\pm 7   $ \\
Mn II     &   2576.877  & $-68\pm 6$   & $249\pm 32$  & $-32\pm 2$  & $318\pm 21 $ \\
          &   2594.499  & $-68\pm 5$   & $73\pm 18$   & $-28\pm 6$  & $478\pm 23 $ \\ 
          &   2606.462  & $-61\pm 3$   & $254\pm 46$  & $-28\pm 3$  & $317\pm 19 $ \\   
Mg II     &   2796.877  & $-81\pm 10$  & $364\pm 12$  & $-20\pm 2$  & $428\pm 30 $ \\
          &   2803.531  & $-80\pm 3$   & $159\pm 31$  & $-19\pm 5$  & $314\pm 28 $ \\  
Mg I      &   2852.963  & $-69\pm 6$   & $373\pm 23$  & $-27\pm 2$  & $167\pm 13 $ \\ 
$^{12}CO$ &   1392.4074 &    -         &    -         & $-23\pm 3$  & $18\pm 4   $ \\
          &   1418.9012 &    -         &    -         & $-26\pm 3$  & $38\pm 12  $ \\ 
          &   1447.2456 &    -         &    -         & $-22\pm 4$  & $28\pm 11  $ \\
          &   1477.4423 &    -         &    -         & $-25\pm 2$  & $50\pm 16  $ \\
\hline 
\end{tabular} 
\end{center}
\label{table:nonlin}
\label{turns}
\end{table*}

\begin{table*}
\caption{Velocity measurements nearby}
\centering
\begin{tabular} {ccc}
\hline
object & galactic   & $<v>_{LSR}$   \\[0.5ex]
       & coordinate & (km/s)        \\ 
	   & (l,b)      &               \\ 
\hline
AR Cas &  112,-2    & -19           \\
       &            &                \\
1 Cas  &  110,+1    & -25            \\
        
\hline
\end{tabular}
\label{table:nonlin}
\end{table*}

\begin{table}
\begin{center} {\footnotesize}
\caption{Column density and depletion}
\begin{tabular}{ccccccc}
\hline
& \multicolumn{3}{c}{$Component 1$} & \multicolumn{3}{c}{$Component 2$} \\
elements(X)   &  ${\rm LogN}$ \footnotemark[1]&  ${\rm LogN}$\footnotemark[2] & $\delta$\footnotemark[3]  & ${\rm 
LogN}$\footnotemark[1] &  ${\rm LogN}$\footnotemark[2] & $\delta$\footnotemark[3] \\
\hline
O         &   15.3       &  -    & -3.16  & 15.2       &  15.2 &   -3.27      \\
C         &   15.5       & 15.5  & -1.44  & 15.2       &  15.2 &   -1.25      \\
Si        &   15.3-16.3  & 15.3  & -1.97  & 15.3-15.9  &  15.3 &   -1.98       \\          
Al        &   14.2       & 13.8  & -2.24  & 14.2       &  14.0 &   -1.95       \\
Fe        &   14.0-14.8  & 14.41 & -2.71  & 14.0-14.9  &  14.0 &   -3.13       \\
Ni        &   13.4       & 14.3  & -5.30  & 14.4       &  13.6 &   -4.61       \\
Mn        &   13.7-14.2  & 14.2  & -0.84  & 14.0-14.2  & 13.35 &   -1.71       \\
Mg        &   14.2-14.5  & 14.25 & -2.97  & 14.0-14.8  & 14.35 &   -2.88       \\
    
\hline 
\end{tabular} 
\end{center}
\label{turns}
\footnote[1]  1 by profile fitting,
\footnote[2]  2 by Curve of growth Method\\
\footnote[3]  3 depletion given by $\delta  = \log (\frac {X}{H})-\log( \frac {X}{H})_{\odot}$\\
\end{table} 

\section{Test of Weaver's model }

Weaver's (1977) model is a straightforward and elegant application of basic physics to the formation of interstellar bubbles. He 
assumed that the bubble started forming immediately as the new-born star's stellar wind interacted with the ambient interstellar 
medium and then traced its evolution. His predictions should therefore be directly comparable with the observations. 
Unfortunately, as discussed below, we find significant discrepancies, as have others, between the predictions of the model and the 
observations.

Although the age of the central star is $ 2\times {\rm 10}^{6}$years (Dawanas et al., 2007), we find that, given its size of 1.5 
pc and our observed expansion velocity of  36 ${\rm km} s^{-1}$, the bubble is only $4\times {\rm 10}^{4}$ years old, consistent 
with that derived by Christopoulou (1995). Similar discrepancies have been noted in other systems, such as in the Large Magellanic 
Cloud where the derived age of the bubbles in N11B ($0.1 - 0.5\times {\rm 10}^{6}$ years) is much less than the 2 - $3 \times {\rm 
10}^6$ year age of the exciting OB association LH10 (Naze et al., 2001).

The physics of bubble formation and expansion are well understood (Weaver 1977) and it is most likely that the discrepancies 
between the model and the observations are due to the inhomogeneous nature of the medium into which the bubble is expanding (Lynds 
and Neil, 1983).

At its heart, the discrepancy in this, and in other bubbles, is that the size of the bubble is much less than would have been 
predicted based on the age of the central star and the observed mass loss rate. In order to explain this, Naze et al. (2001) have 
suggested that either the stellar wind is much less than the accepted value, by as much as a factor of 100, or that the bubble is 
expanding into an inhomogeneous medium. Aprilia \& Dawanas  (2004) has proposed, alternatively, that there are actually two phases 
of bubble expansion. In the first phase bubble expands into surrounding medium and becomes unobservable. In the second phase it 
results in formation of Bubble nebula.

\begin{figure}
\centering
\includegraphics*[width=3.1 in,]{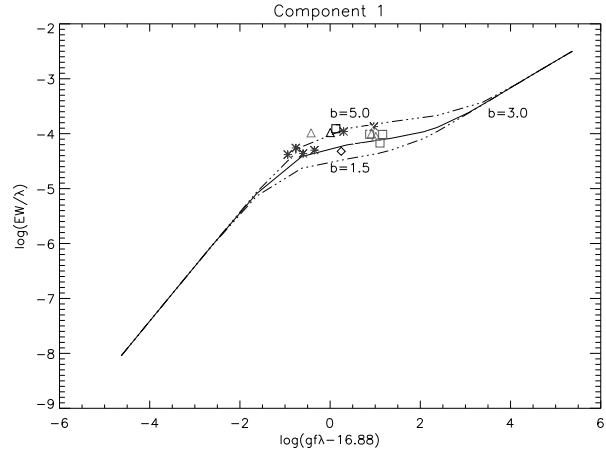}
\caption{Plot of Curve of growth for single ionised elements of component1.}
\end{figure}

\begin{figure}
\centering
\includegraphics*[width=3.1 in,]{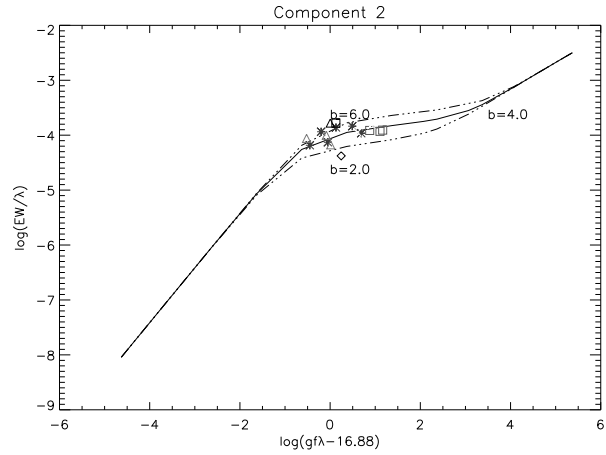}
\caption{Plot of Curve of growth for single ionised elements of component2.}
\end{figure}

\section[]{conclusion } 
We have used IUE high resolution spectra of BD+602522, the central star of the Bubble Nebula, to study the physical properties in
a prototypical interstellar bubble. We have found two components in most of the absorption lines, one from an intervening 
interstellar cloud and the other from the shell around the Bubble. The physical parameters in both components are similar with a 
column density in each of about $10^{21} {\rm cm}^{-2}$ and a temperature on the order of 10000 K, although we are limited by the 
resolution of IUE. We also see evidence for the hot gas in the bubble in the lines of C IV and Al III.

As with other observations of interstellar bubbles, we have found that the derived age of the Bubble Nebula is much less than that 
of the parent star, perhaps indicating that the initial expansion was into an inhomogeneous medium.

There is a rich trove of IUE data on various other bubbles and we are currently investigating these to understand the validity and 
reach of Weaver's original model.

\section*{Acknowledgments}

We thank ISRO for financial support under ISRO-Respond program(project No.10/2/285).

\end{document}